\documentclass{PoS}
\usepackage{amsmath}
\usepackage{graphicx}
\usepackage{psfrag}
\usepackage{cite}

\newcommand{\gfive}{\gamma_5}

\newcommand{\be}{\begin{equation}}
\newcommand{\ee}{\end{equation}}
\newcommand{\bea}{\begin{eqnarray}}
\newcommand{\eea}{\end{eqnarray}}
\newcommand{\ba}{\begin{array}}
\newcommand{\ea}{\end{array}}
\newcommand{\nn}{\nonumber}

\newcommand{\gammaSS}{\gamma_{SS}}

\newcommand{\pref}[1]{(\ref{#1})}

\title{Overlap fermions on a twisted mass sea\footnote{Preprint numbers: DESY 06-160, SFB/CPP-06-41}}

\ShortTitle{Overlap fermions on a twisted mass sea}

\author{O.~B\"{a}r\\ 
        Institute of Physics, Humboldt University Berlin\\
	12489 Berlin, Germany\\
	E-mail: \email{obaer@physik.hu-berlin.de}
}

\author{K.~Jansen, \speaker{S.~Schaefer}, A.~Shindler\\
 NIC, DESY  \\
 Platanenallee 6\\
 15738 Zeuthen, Germany\\
 E-mail: \email{Karl.Jansen,Andrea.Shindler,Stefan.Schaefer@desy.de}
}

\author{L.~Scorzato\\
ECT*\\
Strada delle Tabarelle, 286 \\
38050 Villazzano (TN) Italy  \\
E-mail: \email{scorzato@ect.it}
}

\abstract{
    We present first results of a mixed action project.
    We analyze gauge configurations generated with two flavors of
    dynamical twisted mass fermions. Neuberger's overlap Dirac operator
    is used for the valence sector.
    The various choices in the setup of the simulation are discussed.
    We employ chiral perturbation theory to describe the effects of using
    different actions in the sea and valence sector at non-zero lattice
    spacing.
    }

\FullConference{XXIVth International Symposium on Lattice Field Theory\\
 July 23-28, 2006\\
 Tucson, Arizona, USA}

\begin{document}

\section{Introduction}

Dynamical overlap simulations are extremely time consuming \cite{schaefer}. 
First large scale results have been reported by JLQCD at this conference, but very large computer
resources were needed to obtain them. An efficient alternative to fully dynamical simulations
with chiral fermions might be provided by the so-called mixed action approach.
Here, the fermion discretization employed 
for the sea quarks  is different from the one used for the  valence quarks. 
Provided the sea and valence quark masses are properly matched such a setup is expected to give the same continuum physics as standard unquenched lattice QCD. 

The main idea is to use sea quarks which are (relatively) cheap
to simulate and therefore large volumes, good statistics and 
small quark masses are attainable even with 
the fermion determinant included.
Such fermion discretizations, however, typically break 
chiral symmetry which makes the extraction of certain observables difficult. 
These difficulties can be reduced by using valence quarks that
fulfill the Ginsparg--Wilson equation \cite{Ginsparg:1981bj} and thus 
respect chiral symmetry at finite lattice spacing, e.g. Neuberger's 
overlap fermions \cite{Neuberger:1997fp},  domain wall fermions~\cite{Kaplan:1992bt, Shamir:1993zy}
with a large extent of the fifth dimension or perfect actions~\cite{Hasenfratz:1993sp}.

One drawback of the mixed action approach is that the resulting theory is not
unitary at non-zero lattice spacing. This introduces well known
diseases into the theory
which, however, are expected to vanish in
the continuum limit. Moreover, 
the effects of unitarity violation can be described 
by mixed action chiral perturbation theory~\cite{Bar:2002nr,Bar:2005tu}.

At this conference, we present results obtained on configurations
generated by the European Twisted Mass Collaboration (ETMC) \cite{JansenUrbach,ShindlerICHEP}.
We use an ensemble with two flavors of twisted mass fermions on 
a lattice of size $16^3\times 32$ and a lattice spacing of about $a\approx0.1$~fm.
The (charged) pseudo-scalar mass is about $m_{\rm PS}\approx300$~MeV.

\section{Setup}
The lattices are generated with a tree-level Symanzik improved gauge
action at $\beta=3.9$. Extensive studies have shown that this
gauge action is a good compromise between ease of simulation and 
good short distance behavior \cite{Farchioni:2005ec}.

\subsection{Twisted mass sea fermions}
We take two twisted mass sea fermions into account.
The corresponding Dirac operator reads
\be
D_{{\rm tm}}= D_W^{(N_f=2)}(m_0)  +  i \mu \gamma_5 \tau_3
\ee
with $D_W$ the standard Wilson Dirac operator taken at un-twisted mass
parameter $m_0$. For more details see Refs.~\cite{Frezzotti:2000nk,Frezzotti:2003ni}.
In our simulation we set $\kappa=0.1609$ which 
is close to but not quite the critical value of $\kappa=0.160856$. The twisted 
mass parameter (which determines the bare quark mass) is set to $a\mu=0.004$. 

Twisted mass QCD has many appealing properties. The fermion 
measure is finite and positive everywhere and, if when $\kappa$ is
set to its critical value, physical observables are  automatically
${\cal O}(a)$ improved \cite{Frezzotti:2003ni}.
ETMC has demonstrated that twisted mass fermions can be simulated at small pseudo-scalar masses
and large volume with currently available computer resources, see the talks by
Jansen and Urbach at this conference\cite{JansenUrbach} and Shindler at ICHEP \cite{ShindlerICHEP}.

\subsection{Overlap valence fermions}
We use overlap fermions in the valence sector. They have exact chiral symmetry at finite
lattice spacing.  This ensures ${\cal O}(a)$ improvement at the level of the
action and of the suitably chosen operators. Since also the sea quarks are ${\cal O}(a)$ improved, we
expect small scaling violations for the whole set-up too.
The overlap operator is constructed from the Hermitian Wilson Dirac operator $h=\gfive D_W$
which is taken at a negative mass shift $-R_0$.
\bea
D_{\rm ov}(m)=(R_0-\frac{m}{2})\big[ 1 + \gfive  \epsilon (h(-R_0))   \big] + m
\eea
with $\epsilon$ the matrix sign function and $m$ the bare quark mass.
The parameter $R_0$ is a mass at the cut-off scale which has to be tuned. 
The disadvantage of these fermions is the cost of the construction of
the sign function. This turns out to be ${\cal O}(100)$ times more
expensive than the application of the kernel operator.

\begin{figure}
\begin{center}
 \scalebox{0.75}{
\includegraphics[width=0.55\textwidth,angle=-90]{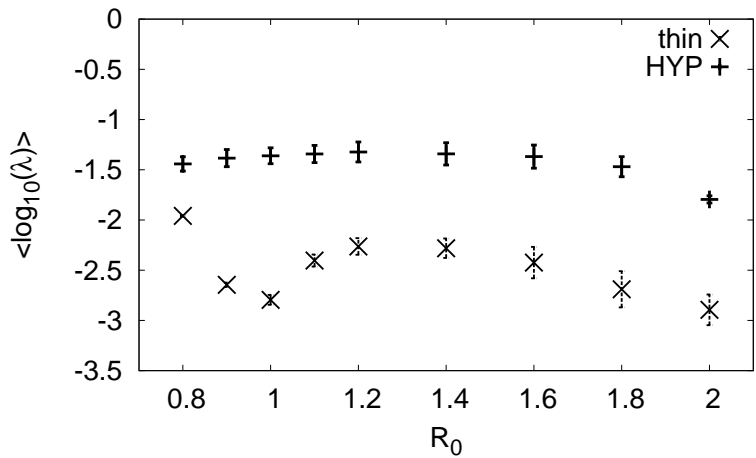}
}
\end{center}
\caption{\label{fig:smearing} The effect of HYP smearing on the low-lying eigenvalues of the
kernel operator $h(-R_0)$. We plot the average of the magnitude of 20th eigenvalue as a function
of the negative mass shift $-R_0$. This value
gives the lower bound of the rational approximation used in the construction of the overlap.
The data is from an ensemble at $\beta=3.75$, $\kappa=0.1661$ and $a\mu=0.005$.}
\end{figure}

We use the standard Hermitian Wilson Dirac operator $h(-R_0)$ as kernel for the overlap,
which we construct  from HYP blocked links\cite{Hasenfratz:2001hp}. 
Among the benefits of the HYP blocking
are lower cost of  construction and better locality of the overlap operator (see Fig.~\ref{fig:smearing}). 
We investigated one and two levels of stout smearing. In particular two steps can lead to
a comparable situation for the overlap operator. However, the smaller range
and  greater (generally positive) experience with the HYP blocking lead us to use 
it in our production runs.

It now remains to determine the best value of the negative mass shift $-R_0$ of the kernel
operator.
We set it such that the resulting overlap operator has optimal locality.
Locality means that the the norm of the operator has an exponential decay at large distances
\[
||D_{\rm ov}(x,y;U)|| \leq C e^{-\nu|x-y|} \ .
\]

\begin{figure}
\begin{center}
 \scalebox{0.8}{
\includegraphics[width=7cm,angle=-90]{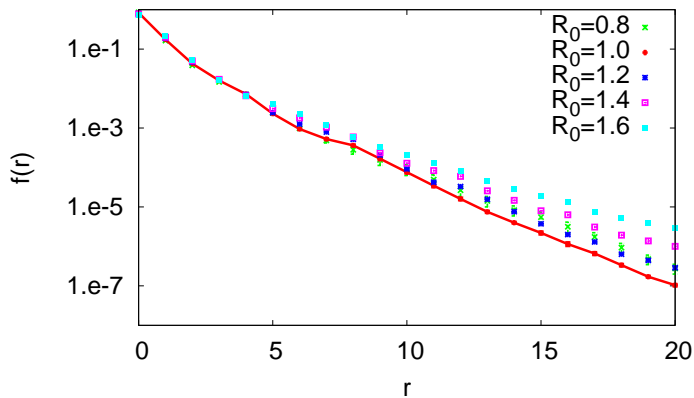}
}
\end{center}
\caption{\label{fig:local}The locality test as defined in Eq.~{\protect \ref{eq:loc}}. We observe
an exponential decay with a decay rate $\nu=0.607$ for the optimal choice $R_0=1$.}
\end{figure}

To verify this, we follow Ref.~\cite{Hernandez:1998et} and measure the exponential decay of $f(r)$
which is essentially the norm of overlap operator on a delta source.
\be
\begin{split}
\psi &= D_{\rm ov} \eta  \ \ {\rm with} \ \ \eta(x)=\delta_{x,x_0}   \\
f(r) &= \mbox{max}\{||\psi(x)||: {\rm dist}(x,x_0)=r \} \label{eq:loc}
\end{split}
\ee
with ${\rm dist}(x,y)$ the taxi-driver distance between $x$ and $y$.
The result is shown in Fig.~\ref{fig:local} for various choices of the parameter
$R_0$. We find that the overlap operator has best locality properties if we set
$R_0=1$, the free field value. (This is a well known effect of the smoothing
of the gauge fields by  HYP blocking.)
For this choice of $R_0$ we measure a decay rate $\nu=0.607$.

\subsection{Matching}
\begin{figure}
\begin{center}
\scalebox{0.8}{
\includegraphics[width=7cm,angle=-90]{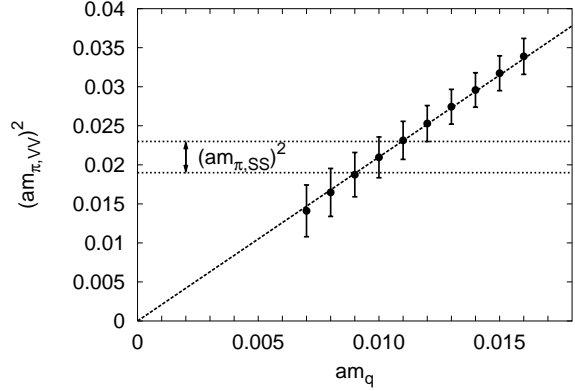}
}
\end{center}
\caption{\label{fig:match}Matching the mass of the charged pseudo-scalar meson
from the valence and the sea quarks. The dashed lines indicate the $1\sigma$
range from the sea sector.}
\end{figure}

The last parameter to fix is the valence quark mass. 
A simple matching condition is given by setting the masses of the pseudo-scalar mesons made of of two sea and two valence quarks equal,
\bea
M^{2}_{SS} & = & M^{2}_{VV}\,.
\eea
For twisted mass fermions, the neutral and the charged pseudo-scalar are not
degenerate. We chose to match the charged meson mass because of  its smaller
statistical uncertainty. In Fig.~\ref{fig:match} we show the valence pseudo-scalar mass
as a function of the bare quark mass. This has to match the meson mass from 
the sea quarks whose $1\sigma$ band is indicated by the dashed lines. The matching
quark mass parameter is close to $am_q=0.01$.

\section{WChPT for twisted mass LQCD}
Mixed action QCD can be described by mixed ChPT provided the pseudoscalars are
sufficiently light \cite{Bar:2002nr}.  Mixed ChPT is formulated analogously to 
ChPT for lattice theories with the same type of sea and valence quarks.
The chiral Lagrangian through O($a^{2}$) for untwisted Wilson sea and GW valence quarks,
which is easily generalized to twisted Wilson sea quarks, can be found in Ref.~\cite{Bar:2003mh}.

At leading order one finds the following expressions for the mesons masses ($m_{V}, \mu_{S}$ denote renormalized quark masses, assuming maximal twist in the sea sector).
\vspace{0.4cm}

\begin{tabular}{lcl}
Val-Val: & \hspace{0.5cm} & $M^{2}_{VV} \,=\, 2Bm_{V}$\\[0.8ex]
Sea-Sea: & \hspace{0.5cm} & $M^{2}_{SS} \,=\, 2B\mu_{S}, \qquad \mu_{S} > \frac{2|c_{2}|a^{2}}{ 2B}$ \\[1.4ex]
Val-Sea: &  \hspace{0.5cm} & $M^{2}_{VS} \,=\, B(m_{V}+\mu_{S}) + a^{2}C_{\rm Mix}$
\end{tabular}
\vspace{0.4cm}

\noindent The valence-valence pion mass vanishes for $m_{V}=0$ because of the exact chiral symmetry in the valence sector. The  minimal sea-sea pion mass $M^{2}_{SS}=2|c_{2}|a^{2}$ is proportional to the low-energy constant $c_{2}$ \cite{Sharpe:1998xm} which determines the phase diagram for the Wilson sea quarks \cite{Munster:2004am,Sharpe:2004ps}. Note that 
\bea
M^{2}_{VS} & \neq &\frac{1}{2}\Big[ M^{2}_{SS} + M_{VV}^{2}\Big]\nn
\eea
because of the extra low-energy constant $C_{\rm Mix}$ present in mixed theories, and a measurement of the mixed meson mass is a direct measure for the size of the O($a^{2}$) cut-off effects in the mixed theory.

Since mixed action theories are not unitary at non-zero lattice spacing, they are 
expected to show the same pathologies as partially quenched theories.
A sensitive probe for these pathologies is the flavor non-singlet scalar correlator
\bea
C(t)& =&  \sum_{\vec{x}}\langle 0| \overline{d}u(\vec{x},t)\, \overline{u}d(\vec{0},0)|0\rangle\nn\,.
\eea
Using the spectral decomposition the scalar correlator can be written as
\bea\label{ScalKorr}
C(t)& =& A e^{-M_{a_0} t} + B(t) + \ldots, 
\eea
where $M_{a_{0}}$ denotes the mass of the $a_{0}$, the lightest $I=1$ scalar meson.
$B(t)$ denotes the contribution of two pseudoscalar states (e.g.\ $\pi\eta^{\prime}$), while the ellipsis 
represents excited scalar states, multi-hadron states, etc.\ which are thought to be less important and therefore ignored.

The contribution $B(t)$ from two pseudoscalar states has been computed in leading order (LO) ChPT including the leading cut-off effects of O($a^{2}$) \cite{Prelovsek:2004jp,Golterman:2005xa}. For large times $t$ one finds approximately ($N_{f}=2$, $M_{SS}=M_{VV}$)\footnote{For simplicity we have sent the singlet mass parameter $m_{0}$ to infinity, 
see Ref.~\cite{Prelovsek:2004jp} for its definition.}
\bea\label{LOBt}
B(t) & \propto & \frac{e^{-2M_{VS} t}}{M_{VS}^{2}} - \frac{e^{-2M_{VV}t}}{M_{VV}^{2}}\Bigg[1 +\frac{\gammaSS a^{2}}{M_{VV}}  t
\Bigg]\,.
\eea
The term proportional to $\gammaSS a^{2}t$ dominates the large $t$ behavior. Note that, depending on the sign of the low-energy constant $\gammaSS$, the scalar correlator can become negative  even if the matching condition $M_{SS}=M_{VV}$ is chosen, in contrast to continuum partially quenched ChPT.

\begin{figure}
\begin{center}
\includegraphics[width=0.35\textwidth,angle=-90]{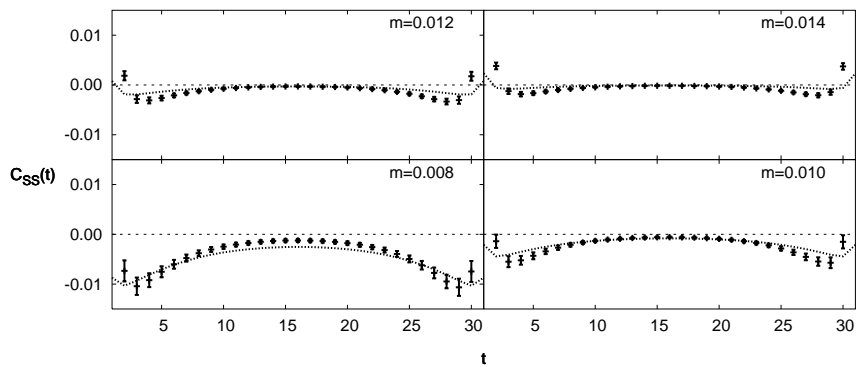}
\end{center}
\caption{\label{fig:sca}The scalar correlator for four values of the valence quark mass. The
dotted lines represent the chiral perturbation theory prediction with the parameters
determined from a fit to the data.}
\end{figure}

Figure \ref{fig:sca} shows the scalar correlator for various valence quark masses, together with a combined fit to the ChPT prediction.\footnote{For the fit we use the exact formula for $C(t)$ with the singlet parameter $m_{0}=1$GeV, not eq.\ \pref{LOBt}.} The scalar correlator is indeed negative for all valence quark masses shown. 
Even though ChPT is able to qualitatively describe the negative correlator, the fit is poor. 
One reason might be that the NLO correction are not negligible.

\section{Summary}
We presented first results from a mixed action project where we use the overlap Dirac operator
in the valence sector and twisted mass fermions in the sea. We discussed the choices of the parameters
of the overlap operator: the smearing of the gauge links for expedience and improved
locality of the operator, the tuning of the radius of the 
Ginsparg--Wilson circle $R_0$ for optimal locality and the determination of the 
valence quark mass such that the pion mass in valence and sea sector match.

As physics result we discussed the scalar correlator and how to describe it with mixed action
chiral perturbation theory. We found good qualitative agreement.
\acknowledgments
This work was supported in part by the DFG Sonderforschungsbereich/Transregio SFB/TR9-03. 
We gratefully acknowledge discussions with T.~DeGrand, N.~Garron, C.~Urbach and U.~Wenger.
We thank T. DeGrand and the MILC collaboration~\cite{MILCcode} on whose codes parts of this work are based.

\providecommand{\href}[2]{#2}\begingroup\raggedright\endgroup

\end{document}